# A METRIC FOR THE ACTIVENESS OF AN OBJECT-ORIENTED COMPONENT LIBRARY

**Sachin Lakra**
Computer Science and Engineering, Manav Rachna College of Engineering, Faridabad, Haryana, India

**Nand Kumar**
Computer Science and Engineering, Manav Rachna College of Engineering, Faridabad, Haryana, India

**Sugandha Hooda**
Computer Science and Engineering, Manav Rachna College of Engineering, Faridabad, Haryana, India

**Nitin Bhardwaj**
Computer Science and Engineering, Manav Rachna College of Engineering, Faridabad, Haryana, India.

*Abstract: In this paper, an attempt has been made to analyze the Activeness of an Object Oriented Component Library (OOCL) and develop a special type of software metric called Component Activeness Quotient(CAQ) which is defined as the degree of readiness of an OOCL. The advantages of the CAQ include a possible comparison between various OOCL's leading to selection of the best OOCL for use during the development task, and Stability of the software can be gauged as indicated by the value of the CAQ. The disadvantage of the CAQ is that it may have some error because of its subjective and random nature.*

*The paper also tries to improvise the calculation of the Activeness Quotient described in [1]. The extreme case of a software organization having an RQ > 1 and MQ=0 was not handled by the method of taking an average of RQ and MQ to calculate the AQ. The improvisation is that the AQ must be equal to a product of MQ and RQ and this is mentioned in the Appendix.*

## 1. Introduction

Object Oriented Programming (OOP) [3, 6] is based on the concept of components being assembled together to create a software. The concept is borrowed from the idea of integrated circuits or chips, resistors, capacitors, etc., being brought together on a printed circuit board to form a piece of hardware. The components in hardware can be picked "off-the-shelf", placed on a printed circuit board and then soldered onto it. Similar is the case with OOP where software components, i.e., predefined class definitions stored in a component library are "picked" and "placed" in a program. The "features" and "actions" of the software components, i.e., their properties and methods, respectively, are then used to implement the proposed software. This is the general approach with OOP.

Programmers need to work very fast on a software project to meet project completion deadlines so as to prevent schedule overruns. Besides this they have to maintain quality so that the software remains stable after completion and major changes are not required. For these two aspects, the software components must be readily available to the programmer in the form of a component library. The component library must be accessible to the programmer from anywhere in the programming environment he is working on.

This paper examines the problem of the "*Activeness*" of a component library and develops a software metric [2,4,5] called the *Component Activeness Quotient (CAQ).* The CAQ represents the readiness of a software component library to make components accessible to a programmer to help him achieve his two-fold goals of preventing schedule overruns and maintaining quality for stability [7] of the software.

The paper is organized as follows: The Introduction is given in Section 1. Section 2 describes the recent concept of Activeness and Section 3 defines an OOCL. Section 4 identifies the steps involved in working in an Object Oriented Environment and the problems involved therein. Section 5 gives the factors involved in measuring the Activeness of an OOCL. Section 6 explains the concept of *Component Activeness Quotient (CAQ)*, its advantages, disadvantage and applications. Concluding remarks are given in Section 7, while Section 8 presents relevant references.

## 2. Activeness

### 2.1 A Recent Concept.

Activeness is a recent new concept which defines the degree of readiness of a system to respond to the stimuli from the environment in which it exists [1]. Activeness may be related to any living entity or any system which exists in any environment in the universe. The Activeness of vacuum without any stimuli and no system existing in it is nil and beyond the scope of our consideration. In a system which shows some activity but the environment does not give any stimuli, Activeness is said to be *closed*. A system which is able to give some response to a stimulus is said to have *open* Activeness [1].

### 2.2 Why does Activeness exist in a system?

The question arises as to why a given system possesses Activeness. The answer is that every system has some "Organizedness (or Capability)" in it, that is, there is some degree of order in the system. The definition of a system itself says that "a System is a set of components working together to achieve a goal." Components cannot work together if they are not organized. This, in turn, implies that if a system is given an external stimulus there will be some change in the degree of order of the system, that is, there will be a response of the system to the stimulus. Every system responds to such stimuli. The "Organizedness" of the system makes it ready to respond to them to some degree. This degree of readiness of a system to respond to a stimulus is the concept called Activeness.

### 2.3 What is the need to study the Activeness of a system?

Another important question is that why Activeness should be studied at all. The answer lies in the fact that the observer wants to know how far she can depend on a system and its response. The system can respond well if it has the readiness to do so and if it is stable. It is stable if it is organized, i.e., if it has order. Thus if the Activeness of a system is known, its stability and how well it can respond to the stimulus it will be given, can be gauged.

## 3. Object Oriented Component Library (OOCL) [8]

A Component Library is a repository of predefined software components which can be "picked off-the-shelf" from the Component Library and "placed" in a program. The major advantage of such components is reusability. The reuse of components leads to the avoidance of reprogramming a particular task from scratch. Such component libraries exist in all current Object Oriented Programming Languages and Integrated Packages and will be found in such Languages and Packages in future as well. Examples of such Languages and Packages include:

- Visual C + +
- Java
- . Net ( runs on Microsoft platforms)
- Mono ( a platform comparable to . Net which runs on Linux, currently under development )

## 4. Working in an Object Oriented Environment (OOE)

Work in an OOE is done in an organized manner based upon the availability of components in OOCL's.

### 4.1 Steps involved in working in an OOE

The steps involved in working in an OOE are: [2]
1. Identify candidate components,
2. Look up components in an OOCL,
3. Extract components if available,
4. Build components if unavailable,
5. Put components in the OOCL, and
6. Use components in a software project.

## 4.2. Problems in working in an Object Oriented Environment (OOE)

The problems in working in an object oriented environment with respect to accessing components from an OOCL include:-
- If the OOCL does not contain the required component, the component may be available at another location, in which case the programmer has to search these other locations. The locations may be local, such as in another OOCL, or remote, i.e., on a server in a network or at a website on the World Wide Web. Accessing such locations requires a large amount of time which leads to the deadline for the software project getting delayed.
- The component may not be available at all in the OOCL or at any other location, which leads to the need to develop this newly required component again adding to the possibility of a schedule overrun.
- The OOCL may not be organized well leading to the waste of time spent in searching for components in the Library.
- The non-availability of the component undermines the overall concept of reusability itself.

## 5. Factors involved in measuring the Activeness of an OOCL.

There are three major factors upon which Activeness of an OOCL is dependent. These include availability of components, access time of the component and the "Organizedness" of the OOCL.

### 5.1 Availability of Components.

The component is either available in the OOCL or it is not available at all. If the component is not available then the Activeness of the OOCL is simply 0 with respect to that component otherwise it is 1.

### 5.2 Access Time of a Component.

Time in which the component can be accessed is dependent on the type of environment through which the component is made available to the programmer working at a workstation. The types of environments are:
- Local environment (on a desktop)- which may be:
  - Command Line Interface – e.g., JDK 1.3
  - Integrated Development Environment with a Graphical User Interface – e.g., VC++
- Remote environments which may be:-
  - A networked environment
  - The Internet

### 5.3 "Organizedness" of an OOCL.

The "organizedness" of an OOCL is dependent on the type of organization of the OOCL which may be:
- Hierarchical
- Search based
- Drop down list within the IDE for which it is necessary for the programmer to know the name of the component she requires.

## 6. The Component Activeness Quotient (CAQ).

**6.1 Def.** *The CAQ may be defined as the degree of readiness of an OOCL.* It may also be defined as the measure of readiness of an OOCL to provide a component to a programmer per unit time and is given by:

$$CAQ = \frac{A_c * R_l}{t} \quad (1)$$

where $A_c$ = Availability of the component in the OOCL(can be 1 or 0),
$R_l$ = Organizedness of the OOCL (a weight according to the type of organization of the OOCL),
t = Access time of the component, in seconds.

The CAQ is dependent on the factors given in section 5. The value of $A_c$ may be 1 or 0 depending upon whether a component is available or it is not available. The value of $R_l$ depends upon the method followed by the OOCL to make components accessible to the programmer. The faster the method, the greater is the weightage of the method and greater is the "organizedness" of the OOCL. The access time t can be obtained experimentally by observing programmers at work.

### 6.2 Advantages

The advantages of the CAQ include:
- Improvement required in the "Organizedness" of the OOCL can be gauged depending upon the value of the CAQ.
- Comparison between OOCL's can be made and the one with the maximum readiness may be chosen for a given project.
- Direct impact of an OOCL on schedule of the software project can be gauged numerically in the form of the CAQ since the faster the programmer can access components the faster the coding can be done and more the chances of meeting the deadline.
- Stability of the software can be gauged as indicated by the value of the CAQ since the software will be more stable if more components are available in the OOCL and if more components can be reused. Higher the value of the CAQ, more stable is the software.

### 6.3 Disadvantage

The CAQ may have some error because it is subjective and stochastic in nature.

### 6.4 Applications of CAQ

The CAQ can be applied in the software industry wherever software development work is conducted in an OOE to gauge the effectiveness of providing an OOCL to a programmer.

## 7. Conclusion

Thus improvements in the OOCL can be made based on the value of the CAQ. The stability, organizedness, availability and responsiveness are features of the OOCL which can be enhanced by taking indications from the CAQ. Further work will be done by the authors of this paper in this direction in the form of a case study.

## 8. References

[1] Sachin Lakra, Bharti Jha, Nitin Bhardwaj, Ritu Saluja and Nand Kumar, "Metrics For The Pre-Development Phase Of Software Requirements Engineering"; Proceedings (Abstract) of National Conference on Emerging Trends in Software Engineering and Information Technology, Gwalior Engineering College, Gwalior, M.P., India; 29-30 March, 2007, pp. 21.

[2] Vaidya Dhananjay and Bhalerao Sidharth, "Overview of the existing Object Oriented Metrics and Frameworks"; Proceedings (Abstract) of National Conference on Emerging Trends in Software Engineering and Information Technology, Gwalior Engineering College, Gwalior, M.P., India; 29-30 March, 2007, pp 15.

[3] Roger S. Pressman, "Software Engineering: A Practitioner's Approach", McGraw Hill, Sixth Edition, 2005.

[4] S. H. Kan, "Software Quality Engineering: Metrics and Models", Pearson Education, Asia, 2002.

[5] N. Fenton and Shari Pfleeger, "Software Metrics", Thomson Asia, Singapore, 2002.

[6] Grady Booch, *Object-Oriented Analysis and Design with Applications*, 2$^{nd}$ Edition, Addison-Wesley.

[7] www.pcmag.com/encyclopedia _term

[8] http://en.wikipedia.org/

# APPENDIX
# METRICS FOR THE PRE-DEVELOPMENT PHASE OF SOFTWARE REQUIREMENTS ENGINEERING [1]

## 1. Requirements Engineering

Requirements analysis is an important part of the software engineering process; whereby business analysts or software developers identify the needs or requirements of a client; having identified these requirements they are then in a position to design a solution.

## 1.1 Requirements Engineering is not sufficient in itself.

The existence of infrastructure requirements also needs to be gauged so as to have some control over the project before it starts – that is, at the pre-development stage.
The infrastructure requirements for a software development project are:-
- Hardware and Software Requirements
- Manpower Requirements

Examples of Hardware Requirements include Workstations, Servers, Backup Power Supply (UPS or other), Equipment for Networking, etc. Software Requirements depend on the domain of the project being developed. Examples of domains include System Software Development, Application Software Development, Website Development and Embedded Software Development. Each of these domains may (or may not) include Software components such as Compilers/Assemblers, Operating Systems, Database Management Systems, Web Servers, CASE Tools, etc.
Examples of Manpower Requirements in the form of skills required in an employee or a person include knowledge of Requirements Engineering, analytical skills, knowledge of modeling tools, knowledge of the programming language, knowledge of testing methodologies and domain knowledge.

## 2. Activeness

Activeness is a new concept which defines the degree of readiness of a system to react to the stimuli from the environment in which it exists. It may be related to any living entity or any system which exists in any environment in the universe. In vacuum, without a system and no stimuli ACTIVENESS = 0.

## 3. Activeness Analysis

Activeness Analysis is the analysis of gauging the *activeness* of an organization, that is, the readiness of an organization to develop a new project. Activeness Analysis may be performed on any software development project at the *pre-development* stage of Requirements Engineering.

## 4. Metrics for Activeness of an organization

- Activeness Quotient (AQ)

- AQ uses two other metrics
  - Resources Quotient (RQ)
  - Manpower Quotient (MQ)

## 4.1 Activeness Quotient (AQ)

Activeness Quotient (AQ) = (RQ * MQ)
where  RQ = Resources Quotient
and    MQ = Manpower Quotient

## 4.2 Resources Quotient (RQ)

$$\text{Resources Quotient (RQ)} = \frac{HS_a}{HS_n} \quad (2)$$

Where  $HS_a$ = Number of Hardware and Software Requirements available
and    $HS_n$ = Number of Hardware and Software Requirements needed

## 4.3 Manpower Quotient (MQ)

$$\text{Manpower Quotient (MQ)} = \frac{[S_a(p_1)/S_r(p_1) + \ldots + S_a(p_n)/S_r(p_n)]}{n} \quad (3)$$

where,  $S_a(p_n)$ = Number of Skills Available in a person $p_n$,
        $S_r(p_n)$ = Number of Skills Needed in a person $p_n$,
and     n = Number of Persons Needed.

## 4.4 Interpretation of AQ

| VALUE OF AQ | INTERPRETATION |
|---|---|
| AQ > 1 | MORE THAN READY |
| AQ = 1 | EXACTLY READY |
| 0 < AQ < 1 | LESS THAN READY – NEED MORE RESOURCES AND/OR MANPOWER |

Table 1 : Interpretation of AQ

## 4.5 Advantages of AQ

- Team Leader can decide how equipped his team is.
- Organizational Head can make a quantified assessment of which new projects can be started with the constraint on resources available.
- A single number, in the form of the Activeness Quotient, can represent the

- readiness of the organization to start work on a project.
- A comparison of the values of the AQ of various software development organizations can allow customers to choose the most "ready" organization.
- The AQ can be converted to a percentage also for convenience by simply multiplying it by 100.

## 4.6 Disadvantages of AQ

The major disadvantage of the AQ is that for a large organization, it may be difficult to obtain a single accurate organizational representative number in the form of the Activeness Quotient if there are projects in the organization which are already in progress, i.e., in various advanced phases of development.

# 5. Conclusions

- Activeness Analysis may be applied to any project.
- Activeness Analysis can be generalized to projects in other fields such as those in the Manufacturing industry and the Research and Development sector.
- Activeness Analysis can be made a part of the Software Requirements Specification since it involves both the customer and the developer.